\documentstyle[prl,aps,multicol]{revtex}
\begin{document}
\draft

\input amssym.def
\input amssym.tex
\begin{multicols}{2}

{\bf Altland and Zirnbauer reply:} In the preceding comment\cite{cis}
Casati, Izrailev and Sokolov (CIS) claim that our recent field
theoretical analysis of the quantum kicked rotor\cite{az} misses an
important dynamical feature: the difference in behavior between
rational and irrational values of $T = \tau/4\pi$ ($\tau$ being the
time between kicks).  We reject that claim.  The rationality of $T$
{\it does} play a fundamental role in our Letter (referred to here as
I).  We were kept from reviewing the number theoretic aspects in I by
space limitations, and welcome this opportunity to dispel any
confusion that may have resulted.

It is well known (cf. Ref.~[10] of I for a review) that the quantum
physics of the rotor is qualitatively different for rational as
opposed to irrational $T$.  In our Letter {\it only the rational case}
was treated. Indeed, what we did was to introduce an upper angular
momentum cutoff $L$ (or, equivalently, we put the system on an angular
lattice), thereby imposing a quantization condition that forces $T$ to
be rational.  To make this point clear, let us briefly review how the
topology of phase space relates to the period $\tau$: Take the
classical phase space $\Gamma$ of the rotor, which is a cylinder ${\rm
S}^1 \times {\Bbb R}$, and compactify it to a 2-torus ${\rm S}^1
\times {\rm S}^1$ by imposing periodic boundary conditions not only
for the angle $\theta$ but also for the angular momentum $l$ (with
period $L \in {\Bbb N}$).  Canonical quantization then requires the
eigenvalues of the operator $\hat\theta$ to be spaced by $2\pi / L$,
thus giving the angular lattice.  Moreover, $L$-periodicity of the
Floquet operator $U$ ($U_{ll'} = U_{l+L,l'+L}$) forces the period
$\tau$ to satisfy a quantum resonance condition, i.e. to be of the
rational form $\tau = 4\pi n / L$ with $n \in {\Bbb N}$.  Conversely,
if $\Gamma = {\rm S}^1 \times {\Bbb R}$ and $\tau$ fulfills the
quantum resonance condition, $U$ commutes with the operator
translating $l$ by $L$ units and Bloch's theorem says that the
eigenfunctions $\psi$ of $U$ organize into sectors labeled by a Bloch
wave number $\alpha$: $\psi(l+L) = e^{i\alpha L} \psi(l)$.  By
choosing to focus on one specific sector, we arrive at the
compactified model with periodic $(\alpha=0)$ or twisted $(\alpha\not=
0)$ boundary conditions.  Thus, the quantum resonance of the kicked
rotor does not ``mysteriously disappear between eqs.~(6) and (8)'' of
I, as is stated by CIS, but is fundamental (though implicit) to the
formulation of the model we treat.

Why did we choose to compactify?  Our main motivation was that
discrete symmetries such as those due to the number theoretic
properties of $T$, are not easy to keep track of in the approximate
steps that are performed in the late stages of our analysis. These
steps are valid if and only if the dominant configurations in (6) are
$Z_{ll'} = \delta_{ll'} Z_l$ where the field $Z_l$ varies slowly with
$l$. In the presence of exact or approximate discrete symmetries, they
become invalid in general.  For example, for $T$ close to rational
$n/L$, where $e^{i\hat\theta L} U e^{-i\hat\theta L} \simeq U$,
additional low-energy modes appear at wave number $\sim 1/L$.  These
are lost by the naive gradient expansion without prior
compactification.  We avoided this difficulty by tuning $T$ to a
rational value and taking care of the resulting discrete symmetry by
restriction of the phase space.  (A similar desymmetrization strategy
was used in order to deal with time reversal symmetry; see footnote
[13] of I.)

Another benefit from compactification is that we {\it can} break time
reversal symmetry by kicking the rotor with $k\cos(\hat\theta + a)$,
where $a\in [0,2\pi/L]$ is closely analogous to an Aharonov-Bohm flux
threading a mesoscopic metallic ring.  CIS argue that $a$ is a pure
gauge, which cannot ever affect the two-particle Green function (1)
of\cite{cis}.  While this is so for $\Gamma = {\rm S}^1 \times {\Bbb
R}$, it is {\it not} true for the cohomologically nontrivial ring
topology we consider\cite{fn1}. 

By construction, the field theory of I inherits the ring topology,
i.e.  the nonlinear $\sigma$ model field $Q$ obeys the boundary
condition $Q(l) = Q(l+L)$.  What are the consequences?  The answer
depends on the size of $L$ relative to $\xi = k^2/2 + ...$, the
localization length.  For $L < \xi$ the quantum motion is diffusive
and states extend more or less uniformly around the ring.  In the
opposite limit $(L \gg \xi)$, diffusion is brought to a halt by
quantum interference effects causing localization.  However, the
wavefunctions are expected to have exponentially small tails, reaching
around the ring even in this case.  True localization is possible only
in the irrational limit $L\to\infty$.  Let us emphasize that the
emergence of localization for $L>\xi$ {\it is not incompatible} with
the resurgence of wavefunctions that is predicted by Bloch's theorem.

To conclude, we reject the insinuation that there is a ``hidden
random-matrix ensemble'' in our Letter.  We are confident that the
steps leading to the field theory (8) are valid for the compactified
or angular lattice model under consideration.  The dominance of
diagonal fields $Z_{ll'} = \delta_{ll'} Z_l$ is ensured by the limit
of hard driving $k\tau \gg 1$, while the gradient expansion is an
expansion in powers of $k/\xi \sim k^{-1}$, valid for large $k$.\\[0.1cm]
Alexander Altland$^{1),2)}$ and Martin R. Zirnbauer$^{1)}$\\[0.3cm]
1) Institut f{\"u}r Theoretische Physik, Z{\"u}lpicher Str. 77, 50937
K{\"o}ln, Germany\\
2) Cavendish Laboratory, Madingley Road, Cambridge CB3 OHE, UK

\end{multicols}

\begin{thebibliography}{9}
\bibitem{fn1} The gauge transformation that eliminates $a$ violates 
the boundary condition $\psi(l+L) = e^{i\alpha L} \psi(l)$, unless 
$a \in 2\pi {\Bbb Z}/L$.
\end{thebibliography}
\end{document}